\documentclass[11pt,a4paper]{article}

\begin{document} 

\begin{flushright}
November, 2007 \\
OCU-PHYS 280 \\
\end{flushright}
\vspace{20mm}

\begin{center}
{\bf\Large
Separability of Dirac equation \\
in higher dimensional Kerr-NUT-de Sitter spacetime}
\end{center}

\begin{center}

\vspace{15mm}

Takeshi Oota$^a$\footnote{
\texttt{toota@sci.osaka-cu.ac.jp}
} and
Yukinori Yasui$^b$\footnote{
\texttt{yasui@sci.osaka-cu.ac.jp}
}

\vspace{10mm}

\textit{
${{}^a}$Osaka City University
Advanced Mathematical Institute (OCAMI)\\
3-3-138 Sugimoto, Sumiyoshi,
Osaka 558-8585, JAPAN
}

\vspace{5mm}

\textit{
${}^b$Department of Mathematics and Physics, Graduate School of Science,\\
Osaka City University\\
3-3-138 Sugimoto, Sumiyoshi,
Osaka 558-8585, JAPAN
}

\vspace{5mm}

\end{center}
\vspace{8mm}

\begin{abstract}
It is shown that the Dirac equations in general higher dimensional Kerr-NUT-de Sitter
spacetimes are separated into ordinary differential equations.\\
\end{abstract}

\vspace{25mm}

\newpage

\renewcommand{\thefootnote}{\arabic{footnote}}
\setcounter{footnote}{0}


Recently, the separability of Klein-Gordon equations in
higher dimensional Kerr-NUT-de Sitter spacetimes \cite{CLP} 
was shown by
Frolov, Krtou\v{s} and Kubiz\v{n}\'{a}k \cite{FKK}.
This separation is deeply related to that of geodesic Hamilton-Jacobi equations.
Indeed, a geometrical object called conformal Killing-Yano tensor plays an important
role in the separability theory \cite{FK,KF,PKVK,FKK,KKPF,KKPV,HOY,HOY2}. 
However, at present, a similar separation of the 
variables of Dirac equations is lacking, although the separability 
in the four dimensional Kerr geometry was given by Chandrasekhar \cite{C}.
In this paper we shall show that  Dirac equations can also be separated in general
Kerr-NUT-de Sitter spacetimes. 

The $D$-dimensional Kerr-NUT-de Sitter metrics are written as follows \cite{CLP}:\\
\noindent
(a) $D=2 n$ 
\begin{equation}\label{EQ1}
g^{(2n)} = \sum_{\mu=1}^{n} \frac{d x_{\mu}^2}{Q_\mu(x)}+
\sum_{\mu=1}^{n} Q_{\mu}(x) \left( \sum_{k=0}^{n-1} A^{(k)}_{\mu}
d \psi_k \right)^2 ,
\end{equation}
\noindent 
(b) $D=2 n+1$
\begin{equation}\label{EQ2}
g^{(2n+1)} = \sum_{\mu=1}^{n} \frac{d x_{\mu}^2}{Q_\mu (x)}+
\sum_{\mu=1}^{n} Q_{\mu}(x) \left( \sum_{k=0}^{n-1} A^{(k)}_{\mu}
d \psi_k \right)^2 
+\frac{c}{A^{(n)}}\left( \sum_{k=0}^{n} A^{(k)}
d \psi_k \right)^2 .
\end{equation}
The functions $Q_{\mu}$ ($\mu=1,2, \cdots , n$) are given by
\begin{equation}
Q_{\mu}(x)=\frac{X_{\mu}}{U_{\mu}},~~~
U_{\mu}=\prod_{\stackrel{\scriptstyle \nu=1}{(\nu \ne \mu)}}^{n}(x_{\mu}^2-x_{\nu}^2),
\end{equation}
where $X_{\mu}$ is a function depending only on the coordinate $x_{\mu}$, and
$A^{(k)}$ and $A^{(k)}_{\mu}$ are the elementary symmetric functions of 
$\{ x_{\nu}^2 \}$ and $\{ x_{\nu}^2 \}_{\nu \neq \mu}$ respectively:
\begin{equation}
\prod_{\nu=1}^{n}(t-x_{\nu}^2)=A^{(0)}t^n-A^{(1)} t^{n-1}+ \cdots + (-1)^n A^{(n)},
\end{equation}
\begin{equation}
\prod_{\stackrel{\scriptstyle \nu=1}{(\nu \ne \mu)}}^{n}(t-x_{\nu}^2)
=A^{(0)}_{\mu}t^{n-1}-A^{(1)}_{\mu}
 t^{n-2}+ \cdots  + (-1)^{n-1} A^{(n-1)}_{\mu}.
\end{equation}
The metrics are Einstein
if $X_{\mu}$ takes the form \cite{CLP,HHOY}\\
\noindent
(a)~$D=2n$
\begin{equation}
X_{\mu}=\sum_{k=0}^{n} c_{2k} x_{\mu}^{2k}+b_{\mu} x_{\mu},
\end{equation}
\noindent
(b)~$D=2n+1$
\begin{equation}
X_{\mu}=\sum_{k=0}^{n} c_{2k} x_{\mu}^{2k}+b_{\mu}
+\frac{(-1)^n c}{x_{\mu}^2},
\end{equation}
where $c, c_{2k}$ and $b_{\mu}$ are free parameters. \\

\noindent
{\bf{1. D=2n }}\\
For the metric (\ref{EQ1}) 
we introduce the following orthonormal basis 
$\{e^a \}=\{e^{\mu},~e^{n+\mu} \}$~($\mu=1,2, \cdots, n$):
\begin{equation} \label{EQ8}
e^\mu=\frac{dx_\mu}{\sqrt{Q_\mu}} ~,
~~~ e^{n+\mu}=\sqrt{Q_\mu}\sum_{k=0}^{n-1}A^{(k)}_{\mu} d\psi_k.
\end{equation}
The dual vector fields 
are given by
\begin{equation} \label{EQ9}
e_{\mu}=\sqrt{Q_{\mu}} \frac{\partial}{\partial x_{\mu}}~,~~~
e_{n+\mu}=\sum_{k=0}^{n-1} \frac{(-1)^k 
x_{\mu}^{2(n-1-k)}}{\sqrt{Q_{\mu}} U_{\mu}}
\frac{\partial}{\partial \psi_{k}}.
\end{equation}
The spin connection is calculated as \cite{HHOY}
\noindent
\begin{eqnarray} \label{EQ10}
\omega_{\mu \nu} &=& -\frac{x_\nu\sqrt{Q_\nu}}{x_\mu^2-x_\nu^2}~e^\mu
-\frac{x_\mu\sqrt{Q_\mu}}{x_\mu^2-x_\nu^2}~e^\nu,~~~(\mu \ne \nu) \nonumber\\
\omega_{\mu, n+\mu} &=& -(\partial_\mu\sqrt{Q_\mu})~e^{n+\mu}
-\sum_{\rho\neq\mu}\frac{x_\mu\sqrt{Q_\rho}}{x_\rho^2-x_\mu^2}~e^{n+\rho},~~~(\mbox{no sum over } \mu),\\
\omega_{\mu, n+\nu} &=& \frac{x_\mu\sqrt{Q_\nu}}{x_\mu^2-x_\nu^2}~e^{n+\mu}
-\frac{x_\mu\sqrt{Q_\mu}}{x_\mu^2-x_\nu^2}~e^{n+\nu},~~~(\mu \ne \nu) \nonumber\\
\omega_{n+\mu, n+\nu} &=& -\frac{x_\mu\sqrt{Q_\nu}}{x_\mu^2-x_\nu^2}~e^\mu
-\frac{x_\nu\sqrt{Q_\mu}}{x_\mu^2-x_\nu^2}~e^\nu, ~~~(\mu \neq \nu). \nonumber
\end{eqnarray}
Then, the Dirac equation is written in the form
\begin{equation}
(\gamma^a D_a + m)\Psi=0,
\end{equation}
where $D_a$ is a covariant differentiation,
\begin{equation} \label{EQ12}
D_a=e_a+\frac{1}{4}\omega_{bc}(e_a)\gamma^b \gamma^c.
\end{equation}
From (\ref{EQ9}),(\ref{EQ10}) and (\ref{EQ12}), 
we obtain the explicit expression for the Dirac equation
\begin{eqnarray} \label{EQ13}
& &\sum_{\mu=1}^{n}\gamma^{\mu}\sqrt{Q_{\mu}}
\Biggl(\frac{\partial}{\partial x_{\mu}}+\frac{1}{2} 
\frac{X_{\mu}^{'}}{X_{\mu}}+
\frac{1}{2}\sum_{\stackrel{\scriptstyle \nu=1}{(\nu \ne \mu)}}^{n} 
\frac{x_{\mu}}{x_{\mu}^2-x_{\nu}^2} \Biggr)\Psi \\
&+& \sum_{\mu=1}^{n}\gamma^{n+\mu}\sqrt{Q_{\mu}}
\Biggl( \sum_{k=0}^{n-1} \frac{(-1)^k x_{\mu}^{2(n-1-k)}}{X_{\mu}}
\frac{\partial}{\partial \psi_k}+\frac{1}{2}
\sum_{\stackrel{\scriptstyle \nu=1}{(\nu \ne \mu)}}^{n} 
\frac{x_{\nu}}{x_{\mu}^2-x_{\nu}^2}(\gamma^{\nu} \gamma^{n+\nu})
\Biggr)\Psi  + m \Psi=0.\nonumber
\end{eqnarray}
Let us use the following representation of $\gamma$-matrices: $\{ \gamma^a, \gamma^b \} = 2 \delta^{ab}$,
\begin{eqnarray} \label{EQ14}
\gamma^{\mu}&=& 
\underbrace{\sigma_3 \otimes \sigma_3 \otimes \cdots \otimes \sigma_3}_{\mu-1}
\otimes \sigma_1 \otimes I \otimes \cdots \otimes I, \\
\gamma^{n+\mu}&=& 
\underbrace{\sigma_3 \otimes \sigma_3 \otimes \cdots \otimes \sigma_3}_{\mu-1} 
\otimes \sigma_2 \otimes I \otimes \cdots \otimes I, \nonumber
\end{eqnarray} 
where $I$ is the $2 \times 2$ identity matrix and $\sigma_i$ are the Pauli matrices.
In this representation, we write the $2^n$ components of the spinor field as
$\Psi_{\epsilon_1 \epsilon_2  \cdots  \epsilon_n }~~(\epsilon_{\mu}=\pm 1)$, and
it follows that
\begin{eqnarray} \label{EQ15}
(\gamma^{\mu}\Psi)_{\epsilon_1 \epsilon_2  \cdots  \epsilon_n }&=& \left(
\prod_{\nu=1}^{\mu-1}\epsilon_{\nu}\right)
\Psi_{\epsilon_1 \cdots  \epsilon_{\mu-1} (-\epsilon_{\mu})
\epsilon_{\mu+1} \cdots \epsilon_n },\\
(\gamma^{n+\mu}\Psi)_{\epsilon_1 \epsilon_2  \cdots  \epsilon_n } 
&=& -i \epsilon_{\mu}
\left(\prod_{\nu=1}^{\mu-1} \epsilon_{\nu}\right)
\Psi_{\epsilon_1 \cdots  \epsilon_{\mu-1} (-\epsilon_{\mu})
\epsilon_{\mu+1} \cdots \epsilon_n }.\nonumber
\end{eqnarray} 
By the isometry the spinor field takes the form
\begin{equation}
\Psi_{\epsilon_1 \epsilon_2  \cdots  \epsilon_n}(x, \psi)= 
\hat{\Psi}_{\epsilon_1 \epsilon_2  \cdots  \epsilon_n}(x) 
\exp \left( i \sum_{k=0}^{n-1} N_k \psi_k \right)
\end{equation}
with arbitrary constants $N_k$. Substituting (\ref{EQ15}) into (\ref{EQ13}), we obtain
\begin{eqnarray} \label{EQ17}
\sum_{\mu=1}^{n} &\sqrt{Q_{\mu}}&\Biggl(
\prod_{\rho=1}^{\mu-1}\epsilon_{\rho}\Biggr)
\Biggl(\frac{\partial}{\partial x_{\mu}}+\frac{1}{2} \frac{X_{\mu}^{'}}{X_{\mu}}+
\frac{1}{2} \frac{\epsilon_{\mu} Y_{\mu}}{X_{\mu}}
+\frac{1}{2}\sum_{\stackrel{\scriptstyle \nu=1}{(\nu \ne \mu)}}^{n}
\frac{1}{x_{\mu}-\epsilon_{\mu} \epsilon_{\nu} x_{\nu}} \Biggr)
\hat{\Psi}_{\epsilon_1 \cdots  \epsilon_{\mu-1} (-\epsilon_{\mu})
\epsilon_{\mu+1} \cdots \epsilon_n }\nonumber\\
&+& m \hat{\Psi}_{\epsilon_1 \epsilon_2  \cdots  \epsilon_n }=0,
\end{eqnarray}
where we have introduced the function
\begin{equation}
Y_{\mu}=\sum_{k=0}^{n-1} (-1)^k x_{\mu}^{2(n-1-k)} N_k,
\end{equation}
which depends only on $x_{\mu}$.

Consider now the region $x_{\mu}-x_{\nu} > 0$ for $\mu < \nu$ and $x_{\mu}+x_{\nu} > 0$.
Let us define
\begin{equation} \label{EQ19}
\Phi_{\epsilon_1 \epsilon_2  \cdots  \epsilon_n }(x)= \prod_{1 \le \mu < \nu \le n}
\frac{1}{\sqrt{x_{\mu}+\epsilon_{\mu} \epsilon_{\nu} x_{\nu}}}. 
\end{equation} 
Then, one can obtain an equality
\begin{equation} \label{EQ20}
\frac{\Phi_{\epsilon_1 \cdots  \epsilon_{\mu-1} (-\epsilon_{\mu})
\epsilon_{\mu+1} \cdots \epsilon_n }(x)}
{\Phi_{\epsilon_1 \epsilon_2  \cdots  \epsilon_n }(x)}=(-\epsilon_{\mu})^{\mu-1}
\Biggl(\prod_{\rho=1}^{\mu-1}\epsilon_{\rho}\Biggr)
\frac{\sqrt{(-1)^{\mu-1} U_{\mu}}}
{\displaystyle \prod_{\stackrel{\scriptstyle \nu=1}{(\nu \ne \mu)}}^{n}
(x_\mu-\epsilon_{\mu} \epsilon_{\nu} x_{\nu} )}.
\end{equation}
Now we show that the Dirac equation allows a separation of variables by setting
\begin{equation} \label{EQ21}
\hat{\Psi}_{\epsilon_1 \epsilon_2  \cdots  \epsilon_n }(x)=
\Phi_{\epsilon_1 \epsilon_2  \cdots  \epsilon_n }(x) \prod_{\mu=1}^{n} 
\chi^{(\mu)}_{\epsilon_{\mu}}(x_{\mu}).
\end{equation}
It should be noticed that
\begin{equation} \label{EQ22}
\frac{\partial}{\partial x_{\mu}} \log 
\hat{\Psi}_{\epsilon_1 \cdots  \epsilon_{\mu-1} (-\epsilon_{\mu})
\epsilon_{\mu+1} \cdots \epsilon_n }=
\frac{d}{d x_{\mu}} \log \chi^{(\mu)}_{-\epsilon_{\mu}}
-\frac{1}{2}\sum_{\stackrel{\scriptstyle \nu=1}{(\nu \ne \mu)}}^{n} \frac{1}{x_{\mu}
-\epsilon_{\mu} \epsilon_{\nu} x_{\nu}}.
\end{equation}
By using (\ref{EQ20}) and (\ref{EQ22}), 
the substitution of (\ref{EQ21}) into (\ref{EQ17}) leads to
\begin{equation}
\sum_{\mu=1}^{n}\frac{P^{(\mu)}_{\epsilon_{\mu}}(x_{\mu})}
{\displaystyle \prod_{\stackrel{\scriptstyle \nu=1}{(\nu \ne \mu)}}^n 
(\epsilon_{\mu} x_{\mu}-\epsilon_{\nu} x_{\nu})}
+m=0,
\end{equation}
where $P^{(\mu)}_{\epsilon_{\mu}}$ is a function of the coordinate $x_{\mu}$ only,
\begin{equation} \label{EQ24}
P^{(\mu)}_{\epsilon_{\mu}}=
(-1)^{\mu-1} (\epsilon_{\mu})^{n-\mu} \sqrt{(-1)^{\mu-1} X_{\mu} } 
\frac{1}{\chi^{(\mu)}_{\epsilon_{\mu}}} 
\left(
\frac{d}{dx_{\mu}}+\frac{1}{2} \frac{X_{\mu}^{'}}{X_{\mu}}
+\frac{\epsilon_{\mu} Y_{\mu}}{X_{\mu}}
\right) \chi^{(\mu)}_{-\epsilon_{\mu}}.
\end{equation}
Putting
\begin{equation}
Q(y)=-m y^{n-1}+\sum_{j=0}^{n-2} q_j y^j
\end{equation}
with arbitrary constants $q_j$, we find
\begin{equation}
P^{(\mu)}_{\epsilon_{\mu}}(x_{\mu})=Q(\epsilon_{\mu} x_{\mu}). 
\end{equation}
Thus, the functions $\chi^{(\mu)}_{\epsilon_{\mu}}$ 
satisfy the ordinary differential equations
\begin{equation} \label{EQ27}
\left(\frac{d}{dx_{\mu}}+\frac{1}{2} \frac{X_{\mu}^{'}}{X_{\mu}}
+\frac{\epsilon_{\mu} Y_{\mu}}{X_{\mu}}
\right) \chi^{(\mu)}_{-\epsilon_{\mu}}-
\frac{(-1)^{\mu-1}(\epsilon_{\mu})^{n-\mu} Q(\epsilon_{\mu} x_{\mu})}
{\sqrt{(-1)^{\mu-1} X_{\mu}}}  \chi^{(\mu)}_{\epsilon_{\mu}}=0.
\end{equation}

\noindent
{\bf{2. D=2n+1 }}\\
For the metric (\ref{EQ2}) we introduce the orthonormal basis 
$\{\hat{e}^a \}=\{\hat{e}^{\mu},~\hat{e}^{n+\mu},~\hat{e}^{2 n+1} \}$~($\mu=1,2, \cdots, n$):
\begin{equation}
\hat{e}^\mu=e^{\mu},
~~~ \hat{e}^{n+\mu}=e^{n+\mu},~~~\hat{e}^{2 n+1}=\sqrt{S}\sum_{k=0}^{n}A^{(k)}d\psi_k
\end{equation}
with $S=c/A^{(n)}$. The 1-forms $e^{\mu}$ and $e^{n+\mu}$ are defined by (\ref{EQ8}). 
The dual vector fields are given by 
\begin{equation}
\hat{e}_{\mu}=e_{\mu}~,~~~
\hat{e}_{n+\mu}=e_{n+\mu}+\frac{(-1)^n}{x_{\mu}^2 \sqrt{Q_{\mu}} U_{\mu}}
\frac{\partial}{\partial \psi_{n}}~,~~~
\hat{e}_{2n+1}=\frac{1}{\sqrt{S}A^{(n)}}\frac{\partial}{\partial \psi_{n}}
\end{equation}
with (\ref{EQ9}). The spin connection is calculated as \cite{HHOY}
\begin{eqnarray}
\hat{\omega}_{\mu \nu}&=& \omega_{\mu \nu},~~~
\hat{\omega}_{\mu, n+\nu}=
\omega_{\mu, n+\nu}+\delta_{\mu \nu} \frac{\sqrt{S}}{x_{\mu}}\hat{e}^{2 n+1},~~~
\hat{\omega}_{n+\mu, n+\nu} = \omega_{n+\mu, n+\nu}, \nonumber\\
\hat{\omega}_{\mu, 2 n+1}&=& \frac{\sqrt{S}}{x_{\mu}}\hat{e}^{n+\mu}-
\frac{\sqrt{Q_{\mu}}}{x_{\mu}}\hat{e}^{2 n+1},~~~
\hat{\omega}_{n+\mu, 2 n+1}= -\frac{\sqrt{S}}{x_{\mu}}\hat{e}^{\mu}. 
\end{eqnarray}
A similar calculation to the even dimensional case yields the following Dirac equation,
\begin{eqnarray}
& &\sum_{\mu=1}^{n}\gamma^{\mu}\sqrt{Q_{\mu}}
\Biggl(\frac{\partial}{\partial x_{\mu}}+\frac{1}{2} \frac{X_{\mu}^{'}}{X_{\mu}}+\frac{1}{2 x_{\mu}}
+\frac{1}{2}\sum_{\stackrel{\scriptstyle \nu=1}{(\nu \ne \mu)}}^{n} \frac{x_{\mu}}{x_{\mu}^2-x_{\nu}^2} 
\Biggr)\Psi \nonumber\\ 
&+& \sum_{\mu=1}^{n}\gamma^{n+\mu}\sqrt{Q_{\mu}}
\Biggl( \sum_{k=0}^{n-1} \frac{(-1)^k x_{\mu}^{2(n-1-k)}}{X_{\mu}}
\frac{\partial}{\partial \psi_k}+\frac{(-1)^n}{x_{\mu}^2 X_{\mu}}\frac{\partial}{\partial \psi_n}
+\frac{1}{2}
\sum_{\stackrel{\scriptstyle \nu=1}{(\nu \ne \mu)}}^{n} 
\frac{x_{\nu}}{x_{\mu}^2-x_{\nu}^2}(\gamma^{\nu} \gamma^{n+\nu})
\Biggr)\Psi\nonumber\\
&+&\gamma^{2n+1} \sqrt{S} \left( -\sum_{\mu=1}^{n} \frac{1}{2 x_{\mu}}
(\gamma^{\mu} \gamma^{n+\mu})+\frac{1}{c}\frac{\partial}{\partial \psi_n} \right) \Psi+ m \Psi=0.
\end{eqnarray}
We use the representation of $\gamma$-matrices given by (\ref{EQ14}) together with
\begin{equation}
\gamma^{2n+1}=\sigma_3 \otimes \sigma_3 \otimes \cdots \otimes \sigma_3.
\end{equation}
Thus, the spinor field $\hat{\Psi}_{\epsilon_1 \epsilon_2  \cdots  \epsilon_n}$ defined by
\begin{equation}
\Psi_{\epsilon_1 \epsilon_2  \cdots  \epsilon_n}(x, \psi)= 
\hat{\Psi}_{\epsilon_1 \epsilon_2  \cdots  \epsilon_n}(x) 
\exp \left( i \sum_{k=0}^{n} N_k \psi_k \right)
\end{equation}
satisfies the equation
\begin{eqnarray} \label{EQ34}
\sum_{\mu=1}^{n} &\sqrt{Q_{\mu}}&\Biggl(
\prod_{\rho=1}^{\mu-1}\epsilon_{\rho}\Biggr)
\Biggl(\frac{\partial}{\partial x_{\mu}}+\frac{1}{2} \frac{X_{\mu}^{'}}{X_{\mu}}+
\frac{1}{2} \frac{\epsilon_{\mu} \hat{Y}_{\mu}}{X_{\mu}} \nonumber\\
& & \qquad \qquad \qquad +  \frac{1}{2 x_{\mu}}
+\frac{1}{2}\sum_{\stackrel{\scriptstyle \nu=1}{(\nu \ne \mu)}}^{n}
\frac{1}{x_{\mu}-\epsilon_{\mu} \epsilon_{\nu} x_{\nu}} \Biggr)
\hat{\Psi}_{\epsilon_1 \cdots  \epsilon_{\mu-1} (-\epsilon_{\mu})
\epsilon_{\mu+1} \cdots \epsilon_n }\nonumber\\
& & + \left( i \sqrt{S}
\Biggl( \prod_{\rho=1}^{n}\epsilon_{\rho}\Biggr)
\left( -\sum_{\mu=1}^{n}\frac{\epsilon_{\mu}}{2 x_{\mu}}+\frac{N_n}{c} \right)+
 m \right)\hat{\Psi}_{\epsilon_1 \epsilon_2  \cdots  \epsilon_n }=0,
\end{eqnarray}
where
\begin{equation}
\hat{Y}_{\mu}=\sum_{k=0}^{n} (-1)^k x_{\mu}^{2(n-1-k)}N_k.
\end{equation}
We find that the Dirac equation above allows a separation of variables
\begin{equation}
\hat{\Psi}_{\epsilon_1 \epsilon_2  \cdots  \epsilon_n}(x)= 
\Phi_{\epsilon_1 \epsilon_2  \cdots  \epsilon_n}(x)
\prod_{\mu=1}^{n} \left( \frac{
\chi^{(\mu)}_{\epsilon_{\mu}}(x_{\mu})}{\sqrt{x_\mu}} \right)
\end{equation}
with $\Phi_{\epsilon_1 \epsilon_2  \cdots  \epsilon_n}$ defined by (\ref{EQ19}).
Indeed, (\ref{EQ34}) becomes
\begin{equation}
\sum_{\mu=1}^{n}\frac{P^{(\mu)}_{\epsilon_{\mu}}(x_{\mu})}
{\displaystyle \prod_{\stackrel{\scriptstyle \nu=1}{( \nu \ne \mu)}}^n 
(\epsilon_{\mu} x_{\mu}-\epsilon_{\nu} x_{\nu})}
+\frac{i \sqrt{c}}{\displaystyle \prod_{\rho=1}^{n}(\epsilon_\rho x_\rho)}
\left( -\sum_{\mu=1}^{n}\frac{\epsilon_{\mu}}{2 x_{\mu}}+\frac{N_n}{c} \right)
+m=0
\end{equation}
with the help of (\ref{EQ24}). Let us introduce the function
\begin{equation}
\hat{Q}(y)=\sum_{j=-2}^{n-1} q_j y^j
\end{equation}
where
\begin{equation}
q_{n-1}=-m,~~~q_{-1}=\frac{i}{2} (-1)^{n-1} \sqrt{c},~~~q_{-2}=\frac{i}{\sqrt{c}}(-1)^n N_n.
\end{equation}
Using the identities
\begin{eqnarray}
\sum_{\mu=1}^{n} \frac{1}{\displaystyle y_{\mu}^2 
\prod_{\stackrel{\scriptstyle \nu}{(\nu \ne \mu)}}(y_{\mu}-y_{\nu})}&=&
\frac{(-1)^{n-1}}{\displaystyle \prod_{\mu=1}^n y_{\mu}} \sum_{\nu=1}^n 
\frac{1}{y_{\nu}} \\
\sum_{\mu=1}^{n} \frac{1}
{\displaystyle y_{\mu} \prod_{\stackrel{\scriptstyle \nu}{(\nu \ne \mu)}}(y_{\mu}-y_{\nu})}&=&
\frac{(-1)^{n-1}}{\displaystyle \prod_{\mu=1}^n y_{\mu}}
\end{eqnarray}
we can confirm that the functions
$\chi^{(\mu)}_{\epsilon_{\mu}}$ satisfy the ordinary differential equations (\ref{EQ27})
by the replacements 
$Y_{\mu} \rightarrow \hat{Y}_{\mu}$ 
and $Q(\epsilon_{\mu} x_{\mu}) \rightarrow \hat{Q}(\epsilon_{\mu} x_{\mu})$.\\

We have shown the separation of variables of Dirac equations
in general Kerr-NUT-de Sitter spacetimes. An interesting problem is to investigate the origin
of separability. In the case of geodesic Hamilton-Jacobi equations and Klein-Gordon
equations we know that the existence of separable coordinates comes from that of a rank-2 
closed conformal Killing-Yano tensor. 
We can also construct the first order differential operators 
from the closed conformal Killing-Yano tensor which commute
with Dirac operators \cite{GRvH,tan,car}.
However, we have no clear answer of the separability of Dirac equations.
As another problem we can study eigenvalues of Dirac operators 
on Sasaki-Einstein manifolds. Indeed, as shown in \cite{CLP,HSY,CLPP1,CLPP2}, the BPS limit
of odd-dimensional Kerr-NUT-de Sitter metrics leads to Sasaki-Einstein metrics.
Especially, the five-dimensional metrics are important from the point of view of AdS/CFT
correspondence.

\vspace{5mm}

\noindent
{\bf{Acknowledgements}}

\vspace{3mm}

We thank G.W. Gibbons for attracting our attention to the references \cite{GRvH,tan,car}.
This work is supported by the 21 COE program
``Construction of wide-angle mathematical basis focused on knots".
The work of Y.Y. is supported by the Grant-in Aid for Scientific
Research (No. 19540304 and No. 19540098)
from Japan Ministry of Education. 
The work of T.O. is supported by the Grant-in Aid for Scientific
Research (No. 18540285 and No. 19540304)
from Japan Ministry of Education.



\baselineskip 5mm
\begin{thebibliography}{99}

\bibitem{CLP}
W. Chen, H. L\"{u} and C.N. Pope, 
``General Kerr-NUT-AdS metrics in all dimensions,''
Class. Quant. Grav. \textbf{23} (2006) 5323-5340,
\texttt{arXiv:hep-th/0604125}.

\bibitem{FKK}
V.P. Frolov, P. Krtou\v{s} and D. Kubiz\v{n}\'{a}k,
``Separability of Hamilton-Jacobi and Klein-Gordon Equations
in General Kerr-NUT-AdS Spacetimes,''
JHEP \textbf{0702} (2007) 005,
\texttt{arXiv:hep-th/0611245}.

\bibitem{FK}
V.P. Frolov and D. Kubiz\v{n}\'{a}k,
```Hidden' Symmetries of Higher Dimensional Rotating Black Holes,''
Phys. Rev. Lett. \textbf{98} (2007) 11101,
\texttt{arXiv:gr-qc/0605058}.

\bibitem{KF}
D. Kubiz\v{n}\'{a}k and V.P. Frolov,
``Hidden Symmetry of Higher Dimensional Kerr-NUT-AdS Spacetimes,''
Class. Quant. Grav. \textbf{24} (2007) F1-F6,
\texttt{arXiv:gr-qc/0610144}.

\bibitem{PKVK}
D.N. Page, D. Kubiz\v{n}\'{a}k, M. Vasudevan and P. Krtou\v{s},
``Complete Integrability of Geodesic Motion in General Kerr-NUT-AdS Spacetimes,''
Phys. Rev. Lett. \textbf{98} (2007) 061102,
\texttt{arXiv:hep-th/0611083}.


\bibitem{KKPF}
P. Krtou\v{s}, D. Kubiz\v{n}\'{a}k, D.N. Page and V.P. Frolov,
``Killing-Yano Tensors, Rank-2 Killing Tensors,
and Conserved Quantities in Higher Dimensions,''
JHEP \textbf{0702} (2007) 004,
\texttt{arXiv:hep-th/0612029}.

\bibitem{KKPV}
P. Krtou\v{s}, D. Kubiz\v{n}\'{a}k, D.N. Page and M. Vasudevan,
``Constants of Geodesic Motion in Higher-Dimensional Black-Hole Spacetime,''
\texttt{arXiv:hep-th/0707.0001}.

\bibitem{HOY}
T. Houri, T. Oota and Y. Yasui,
``Closed conformal Killing-Yano tensor and geodesic integrability,"
\texttt{arXiv:hep-th/0707.4039}.

\bibitem{HOY2}
T. Houri, T. Oota and Y. Yasui,
``Closed conformal Killing-Yano tensor and Kerr-NUT-de Sitter spacetime uniqueness,"
Phys. Lett. \textbf{B656} (2007) 214-216,
\texttt{arXiv:hep-th/0708.1368}.

\bibitem{C}
S. Chandrasekhar,
``The solution of Dirac's equation in Kerr geometry,"
Proc. R. Soc. London \textbf{A349} (1976) 571-575.


\bibitem{HHOY}
N. Hamamoto, T. Houri, T. Oota and Y. Yasui,
``Kerr-NUT-de Sitter curvature in all dimensions,''
J. Phys. \textbf{A40} (2007) F177-F184,
\texttt{arXiv:hep-th/0611285}.

\bibitem{GRvH}
G.W. Gibbons, R.H. Rietdijk and J.W. van Holten,
``SUSY in the sky,"
Nucl. Phys. \textbf{B404} (1993) 42-64,
\texttt{arXiv:hep-th/9303112}.

\bibitem{tan}
M. Tanimoto,
``The Role of Killing-Yano tensors in supersymmetric mechanics on a curved manifold,"
Nucl. Phys. \textbf{B442} (1995) 549-562,
\texttt{arXiv:gr-qc/9501006}.

\bibitem{car}
M. Cariglia,
``Quantum Mechanics of Yano tensors: Dirac equation in curved spacetime,"
Class. Quant. Grav. \textbf{21} (2004) 1051-1078,
\texttt{arXiv:hep-th/0305153}.


\bibitem{HSY}
Y. Hashimoto, M. Sakaguchi and Y. Yasui,
``Sasaki-Einstein Twist of Kerr-AdS Black Holes,"
Phys. Lett. {\bf{B 600}} (2004) 270-274,
\texttt{arXiv:hep-th/0407114}.

\bibitem{CLPP1}
M. Cveti\v{c}, H. L\"{u}, D.N. Page and C.N. Pope,
``New Einstein-Sasaki spaces in five and higher dimensions,"
Phys. Rev. Lett. {\bf{95}} (2005) 071101,
\texttt{arXiv:hep-th/0504225}.

\bibitem{CLPP2}
M. Cveti\v{c}, H. L\"{u}, D.N. Page and C.N. Pope,
``New Einstein-Sasaki and Einstein spaces from Kerr-de Sitter,"
\texttt{arXiv:hep-th/0505223}.

    
\end{thebibliography}
\end{document}